\documentclass[prb, reprint, superscriptaddress, twocolumn]{revtex4}
\usepackage{graphicx}
\usepackage{dcolumn}
\usepackage{bm}
\usepackage{natbib}
\usepackage[colorlinks=true, linkcolor=red, citecolor=blue, urlcolor=blue, linktoc=page, bookmarks=false, pdfstartview={FitH}, pdfborder={0 0 0.0 [3 3]}]{hyperref} 

\begin{document}

\title{Magnetic properties of Mn-doped Ge$_{46}$ and Ba$_8$Ge$_{46}$ clathrates}
\author{Nirmal Ganguli}
\altaffiliation[Also at ]{Department of Physics, Indian Institute of Technology Bombay, Mumbai 400076, India.}
\affiliation{Department of Solid State Physics, Indian Association for the Cultivation of Science, Jadavpur, Kolkata 700032, India.}
\author{K. V. Shanavas}
\altaffiliation[Present address: ]{Department of Physics, University of Missouri, Columbia, Missouri 65211, USA}
\affiliation{High Pressure and Synchrotron Radiation Physics Division, BARC, Mumbai 400085, India.}
\author{Indra Dasgupta}
\altaffiliation[Also at ]{Center for Advanced Materials, Indian Association for the Cultivation of Science, Jadavpur, Kolkata 700032, India}
\email[Email: ]{sspid@iacs.res.in}
\affiliation{Department of Solid State Physics, Indian Association for the Cultivation of Science, Jadavpur, Kolkata 700032, India.}


\begin{abstract}
We present a detailed study of the magnetic properties of unique cluster assembled solids namely Mn doped Ge$_{46}$ and Ba$_8$Ge$_{46}$ clathrates using density functional theory. We find that ferromagnetic (FM) ground states may be realized in both the compounds when doped with Mn. In Mn$_2$Ge$_{44}$, ferromagnetism is driven by hybridization induced negative exchange splitting, a generic mechanism operating in many diluted magnetic semiconductors. However, for Mn-doped Ba$_8$Ge$_{46}$ clathrates incorporation of conduction electrons via Ba encapsulation results in RKKY-like magnetic interactions between the Mn ions. We show that our results are consistent with the major experimental observations for this system.
\end{abstract}
\maketitle

Cluster assembled solids formed by large cages of Si and Ge, the so called clathrates have drawn considerable attention in the recent years because of the wealth of physical properties displayed by them. These periodic arrangement of nano-sized cages may be semiconducting with energy gap larger in comparison to that of the standard diamond form of Si and Ge.\cite{adams94} Alkali/alkaline earth encapsulated Si and Ge clathrates are metals.\cite{moriguchi00} Some of them are found to be superconducting\cite{yuan04} and are also suggested to be excellent candidate for thermoelectric applications.\cite{nasir09} The germanium clathrates were also found to be ferromagnetic upon doping with transition metals (TMs).\cite{kawaguchi00, li03} Ferromagnetism was also reported for Eu$_4$Sr$_4$Ga$_{16}$Ge$_{30}$\cite{woods06} and Eu$_8$Ga$_{16}$Ge$_{30}$\cite{phan08} clathrates. The discovery of magnetism in Ge-based clathrates have added a novel functionality to these cluster-assembled solids that may find possible application in magnetic devices.

\begin{figure}
\centering
\includegraphics[scale = 0.47]{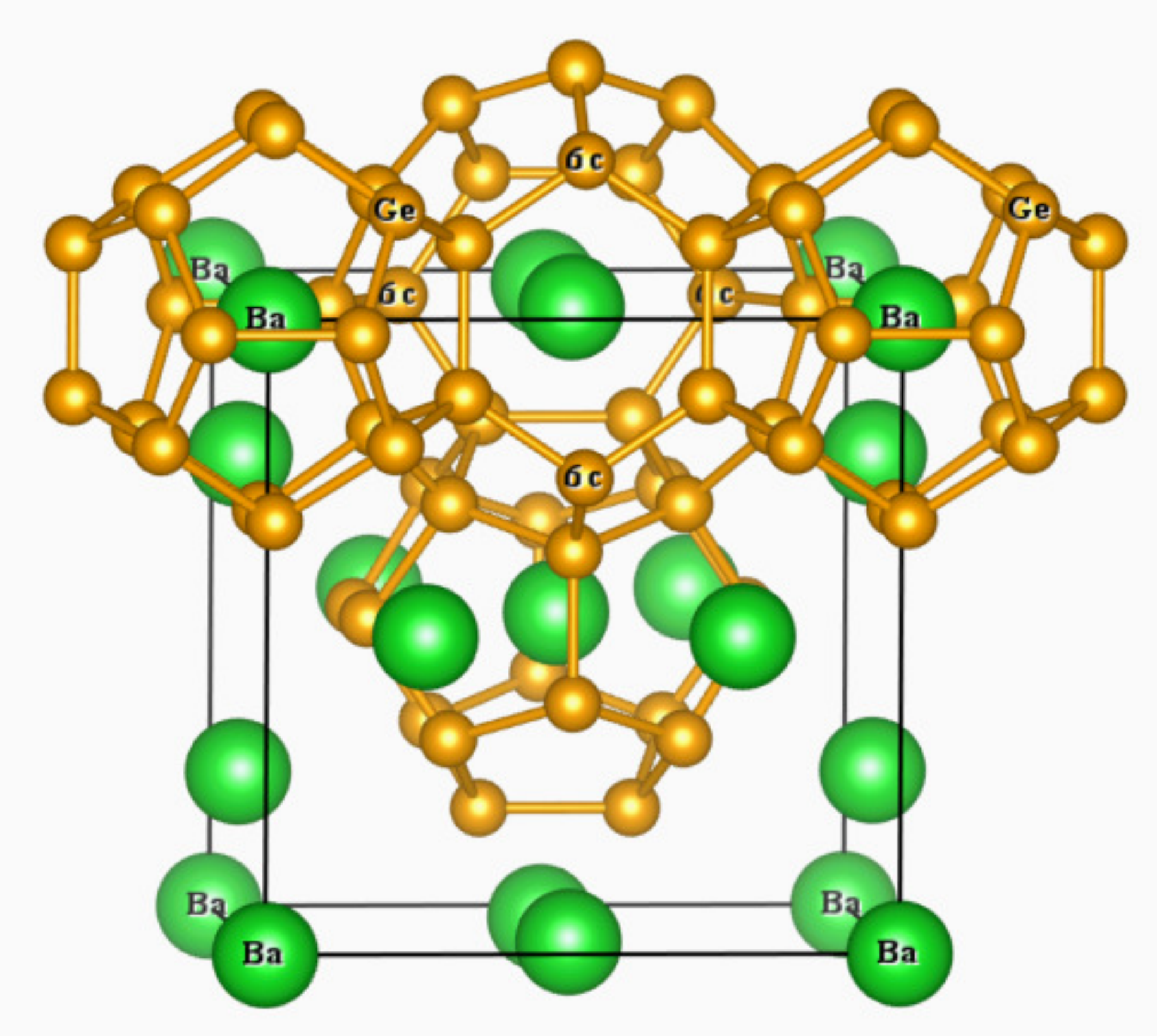}
\caption{\label{fig:structure}(Color online) Structure of Ba$_8$Ge$_{46}$ clathrate showing the simple cubic lattice with Ge$_{20}$ and Ge$_{24}$ cages. Ge atoms at the crystallographic 6c sites are highlighted. All the Ge atoms in the unit cell are not shown for clarity.}
\end{figure}

Ge$_{46}$ is a type~I clathrate with simple cubic lattice, where the Ge atoms form closed cage-like structures of Ge$_{20}$ and Ge$_{24}$. The unit cell consists of two 20 atom (Ge$_{20}$) cages and six 24 atom (Ge$_{24}$) cages. Among the two Ge$_{20}$ cages one is placed at the corner of the cube, while the other, rotated 90$^{\circ}$ with respect to the first is placed at the body center of the cube. Each Ge in Ge$_{20}$ cage is bonded to three other atoms within the same cage. The four fold coordination is completed in two steps. The eight Ge atoms in each Ge$_{20}$ cage form bonds along the eight $\langle 111 \rangle$ directions with the other Ge$_{20}$ unit and the co-ordination of the remaining 12 atoms in each Ge$_{20}$ cage is taken care by adding six Ge atoms at the interstitial 6c sites in the primitive unit cell. The resulting primitive unit cell has 46 Ge atoms (space group $Pm\bar{3}n$) where each Ge is tetrahedrally bonded and therefore the material is expected to be semiconducting. Although semiconducting pure Ge clathrates can be experimentally synthesized,\cite{guloy06} they however prefer to encapsulate alkali and alkaline earth metals like Na, K, and Ba, which occupy the centers of the Ge$_{20}$ and Ge$_{24}$ cages (see Fig.~\ref{fig:structure}). Such encapsulation induces a transition from semiconducting phase to a metallic phase.\cite{zhao99} 

The interstitial 6c sites (highlighted in Fig. 1) are of particular importance as the doping of Mn atoms randomly at these 6c sites of Ba$_8$Ge$_{46}$ clathrate was reported to yield ferromagnetism with saturation magnetic moment per Mn atom to be $\sim$0.8~$\mu_B$ and the ferromagnetic Curie temperature to be $\sim$10~K.\cite{kawaguchi00} The distance between the dopant Mn atoms at the 6c sites was also found to be  quite large,\cite{kawaguchi00} so it was speculated that the interaction between the dopants may be Ruderman-Kittel-Kasuya-Yosida (RKKY)-like. Following this report, Yang~{\em et~al.}\cite{yang04} have investigated this system in the framework of {\em ab~initio} density functional calculations. For a pair of Mn atoms to be substituted at the 6c sites in the unit cell, if the position of one Mn is fixed then the other can occupy any of the remaining five sites. Out of these five neighbors of Mn, one is at a distance $a/2$(configuration~I) while the other four  are at a distance $\sqrt 6 a/4$(configuration~II), where $a$ is the lattice constant. The calculation by Yang et al. in the framework of LDA yielded the value of the magnetic moment per Mn atom to be 0.77~$\mu_B$ and 0.42~$\mu_B$ when the two Mn atoms are doped in configuration~I and configuration~II respectively\cite{yang04}. Although the former value of the magnetic moment agrees well with the experiment but the latter is found to be substantially less. Further, Yang et al. \cite{yang04} did not explore the possibility of antiferromagnetic (AFM) coupling between a pair of Mn atoms necessary for RKKY like interaction. In view of the above, in this letter we have studied the magnetic properties of Mn doped Ba$_8$Ge$_{46}$ and pristine Ge$_{46}$ clathrates in some details using {\em ab~initio} density functional calculations.  Our calculations reveal that ferromagnetism may be realized in both the systems but the origin of ferromagnetism in guest-free Mn$_2$Ge$_{44}$ system is markedly different from metal encapsulated Ba$_8$Mn$_2$Ge$_{44}$ clathrates.

The electronic structure and total energy calculations presented in this paper are performed using {\em ab~initio} density functional theory (DFT) as implemented in the Vienna Ab-initio Simulation Package (VASP).\cite{vasp2, vasp1} The electron-ion interaction in the core and valence part are treated within the projector augmented wave (PAW) method\cite{paw} along with plane wave basis set. We have employed generalized gradient approximation (GGA) due to Perdue-Burke-Ernzerhof (PBE)\cite{pbe} to treat the exchange and correlation.
The localized Mn d-states are treated in the framework of GGA+U method,\cite{dudarevPRB98} where calculations are done for several values of U in the range $2-7$ eV and J=1 eV. Atomic positions were relaxed to minimize the Hellman-Feynman forces on each atom with the tolerance value of 10$^{-2}$~eV/\AA. The optimum values of the energy cutoff and the size of the $k$-point mesh is found to be 650~eV and $4 \times 4 \times 4$ respectively and was accordingly employed in our calculation.

\begin{figure}
\centering
\includegraphics[scale = 0.45]{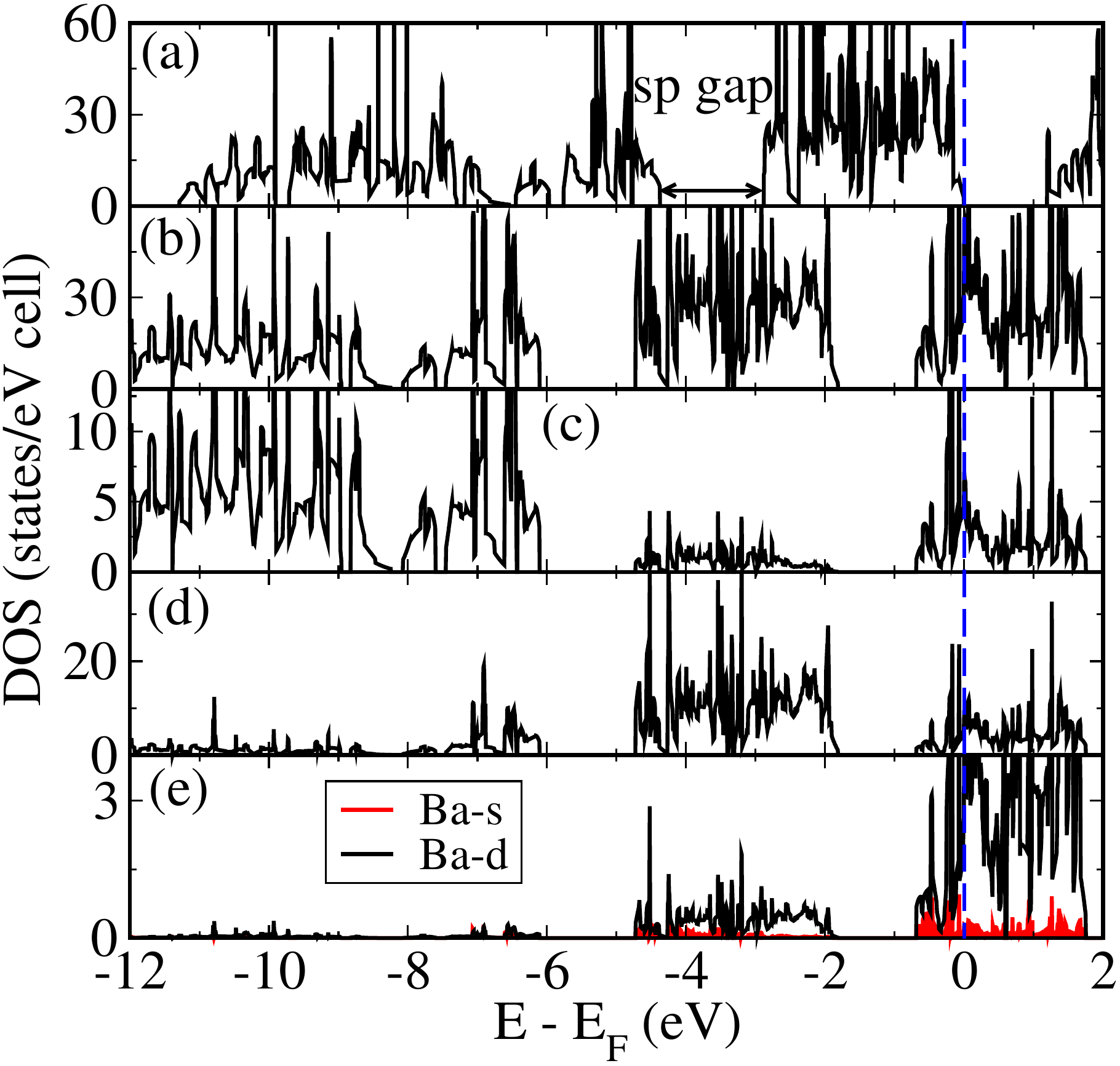}
\caption{\label{fig:Ba8Ge46dos}(Color online) The density of states corresponding to Ge$_{46}$ and  Ba$_8$Ge$_{46}$ clathrates. Subfigure (a) and (b) show the total density of states for Ge$_{46}$ and Ba$_8$Ge$_{46}$ clathrates respectively. Ge-$s$,  Ge-$p$ and Ba-$s$ + Ba-$d$ states corresponding to the Ba$_8$Ge$_{46}$ clathrate are depicted in subfigures (c), (d) and (e) respectively.}
\end{figure}

To begin with we have computed the electronic structure of pristine Ge$_{46}$ and Ba$_8$Ge$_{46}$ clathrates without doping. Our optimized lattice constants for Ge$_{46}$ and Ba$_8$Ge$_{46}$ clathrates are 10.78~\AA\ and 11.01~\AA\ respectively. The density of states(DOS) corresponding to these clathrates are displayed in Fig.~\ref{fig:Ba8Ge46dos}.
We see from Fig.~\ref{fig:Ba8Ge46dos}(a) that Ge$_{46}$ is a semiconductor with a band gap of 1.19~eV. The valence band DOS of the pristine Ge$_{46}$ (Fig.~\ref{fig:Ba8Ge46dos}(a), from left to right) exhibits three major parts as reported earlier\cite{dong99} that may be assigned to a $s$-like region, a $sp$-hybridized region, and a $p$-like region, with a characteristic $s$-$p$ gap. The origin of this $s$-$p$ gap has been attributed to the five ring patterns (pentagons) of the Ge atoms.\cite{SaitoPRB95} We have displayed the total DOS as well as the partial density of states (pDOS) for Ba encapsulated Ge$_{46}$ clathrate in Fig.~\ref{fig:Ba8Ge46dos}(b) and Fig.~\ref{fig:Ba8Ge46dos}(c), (d), and (e) respectively. Upon Ba encapsulation the additional 16 valence electrons are accommodated in the conduction band and the system becomes metallic. While the DOS of the valence band of Ba$_8$Ge$_{46}$ is very similar to pristine Ge$_{46}$, the conduction band DOS shows modification upon the inclusion of metal atoms. The pDOS plot of Ba$_8$Ge$_{46}$ clathrates (see Fig.~\ref{fig:Ba8Ge46dos}(c), \ref{fig:Ba8Ge46dos}(d), and \ref{fig:Ba8Ge46dos}(e)) reveal that the conduction band has considerable admixture of Ge-$s$ and $p$ states with Ba-$s$ and Ba-$d$ states with substantial contribution from the latter. These electrons are expected to be delocalized in the entire system and act as a conduction electron cloud.

Next we have considered doping of a pair of Mn atoms at the 6c sites of the semiconducting pristine Ge$_{46}$ clathrate.  
\begin{figure}
\includegraphics[scale = 0.48]{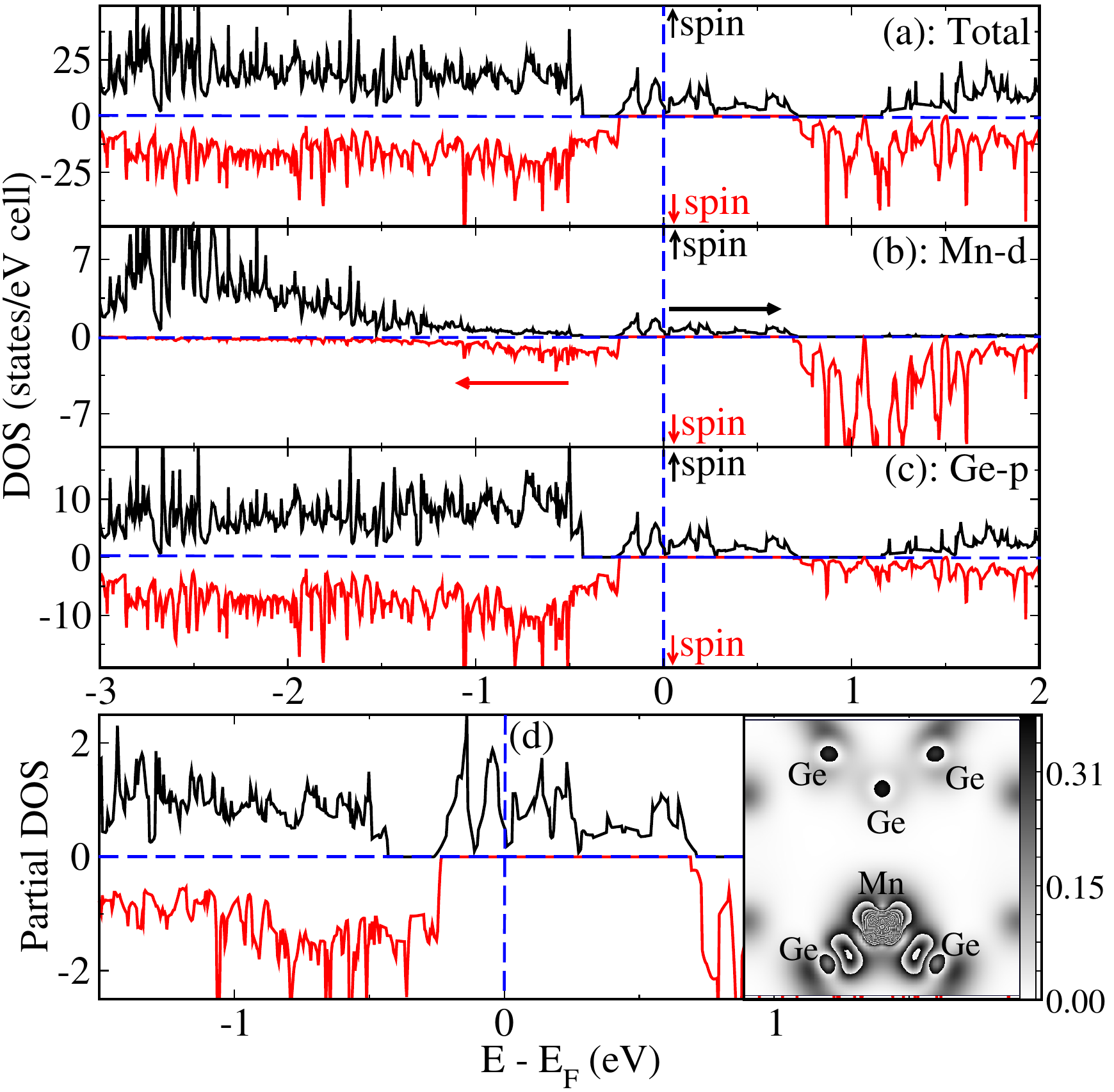}
\caption{\label{fig:noBaDos}(Color online) The spin polarized GGA+U (U=4 eV and J=1 eV) density of states for Mn$_2$Ge$_{44}$, with parallel alignment of Mn spins (FM) in configuration~I. Subfigure (a) shows the total DOS, while subfigures (b) and (c) indicate the projected DOS corresponding to Mn-$3d$ and Ge-$4p$ states. Subfigure (d) represents the partial density of states corresponding to {\em one} dopant Mn atom and its surrounding Ge atoms, whereas the inset shows 2-dimensional charge density for the up spin valence band hole states.}   
\end{figure}
In order to understand the magnetic properties of the doped system we have calculated the magnetic exchange interactions as the energy difference between antiparallel (AFM) and parallel (FM) arrangements of the magnetic moments of a pair of Mn ions for configurations~I and II for several values of U ranging from U=0 (GGA) to U=7 eV. Our calculations reveal that the exchange interaction is always ferromagnetic in both the configurations for all values of U except for U=7.0 eV (see table SI-1).\cite{SI} These results suggest that Mn doped  Ge$_{46}$ may be ferromagnetic. For further insights on the origin of magnetic ordering in ${\rm Mn_2Ge_{44}}$ we have displayed in Fig.~\ref{fig:noBaDos} the spin polarized density of states (DOS) for ${\rm Mn_2Ge_{44}}$ with parallel alignment of the Mn spins (FM) in configurations~I for the representative case, U=4 eV and J=1 eV. The DOS indicates that the system is a half-metallic ferromagnet, analogous to diluted magnetic semiconducting (DMS) materials like Mn-doped GaAs. When a Ge is substituted by a Mn having seven valence electrons, four electrons are utilized to saturate the Ge dangling bonds and the remaining three electrons are responsible for the magnetism. A comparison of the total DOS (Fig.~\ref{fig:noBaDos}(a)) and the pDOS for Mn-$d$ (Fig.~\ref{fig:noBaDos}(b)) and Ge-$p$ (Fig.~\ref{fig:noBaDos}(c)) suggests that the Mn-$d$ states are not only strongly exchange split but also  Mn-$d$ spin up states are fully occupied and the Mn-$d$ spin down states are completely empty confirming that Mn is in high spin $d^5$ configuration. So the introduction of Mn in pristine Ge$_{46}$ gives rise to an acceptor in the Ge -$p$ manifold and the hole produced by the acceptor (see Fig.~\ref{fig:noBaDos}(b)) is expected to interact with the localized Mn-$d$ states to mediate ferromagnetism.
We find from Fig.~\ref{fig:noBaDos}(b) that the exchange split  Mn 3d states in the ferromagnetic configuration strongly hybridizes with the Ge-$p$ states. As a result, the spin up Ge-$p$ bands shift to higher energy while the spin down Ge-$p$ bands make an opposite shift as indicated by the arrows in Fig.~\ref{fig:noBaDos}(b) and therefore an energy gain is obtained by transferring electrons from the Ge spin up states to the Ge spin down ones. The exchange splitting of the $p$-states induced by this hybridization are opposite to that of the Mn-$d$ states (i.e.\ hybridization induced negative exchange splitting) resulting in the calculated magnetic moment of Mn and Ge oppositely aligned and the total magnetic moment to be $6~\mu_B$ per unit cell. This generic mechanism also responsible for ferromagnetism in Sr$_2$FeMoO$_6$\cite{sarma00, fangPRB01, kanamoriJPSJ01} and Mn doped GaAs\cite{mahadevan04} is further corroborated by the plot of pDOS (Fig 3(d)) and charge density  of the valence band hole states in the spin up channel, shown in Fig.~\ref{fig:noBaDos}(d) (inset). From these plots we gather that there is strong hybridization of the Mn$-d$ states with the Ge$-p$ states, thereby accounting for ferromagnetism in Mn doped Ge$_{46}$.
In the presence of strong hybridization between the Mn-$d$ and Ge-$p$ states this novel mechanism of ferromagnetism underscores the competing antiferromagnetic ordering of Mn spins via super exchange.

Finally we have considered the doping of Mn in Ba$_8$Ge$_{46}$. We have first assumed substitutional doping of two Mn atoms into the Ba$_8$Ge$_{46}$ clathrate at the 6c sites. 
The equilibrium lattice constant and the atomic positions are calculated for the Mn doped at the 6c sites  both in configuration~I and configuration~II for the ferromagnetic and antiferromagnetic alignment of Mn spins.The equilibrium lattice constant is calculated to be 11.08~\AA, which is an overestimation from the experimentally reported value by 3.6~\% possibly due to the use of GGA for the exchange-correlation.\cite{dong99}
We have also calculated the exchange interaction and the magnetic moment per unit cell (in the FM state) in configuration~I and configuration~II for values of U ranging from U=0 (GGA) to U=7 eV (see Table~SI-2).\cite{SI} We find that in configuration~I the exchange interaction is antiferromagnetic while in  the configuration~II, it is ferromagnetic, independent of the chosen  values of U (see Table~SI-2).\cite{SI} This is in sharp contrast to the case of Mn doped Ge$_{46}$.  The change of the magnitude as well as the sign of the exchange interaction with distance points to the fact that the RKKY model may be applicable here. It is well known that for a fixed concentration of the magnetic ions, the RKKY interaction depends only on the distance between the magnetic ions in the presence of the delocalized conduction electron cloud. So in order to extract the {\em characteristic} distance dependence of the RKKY interaction, we have also considered doping configurations, where one Mn is at the 6c site but the second one  is in other possible framework sites as the conduction electron cloud spreading the entire clathrate network mediating the exchange interaction will possibly not distinguish among the different framework sites. The results of our calculations for the exchange interaction for the representative case U=4.0eV and J=1.0eV are summarized in Fig.~4.  
\begin{figure}
\includegraphics[scale = 0.43]{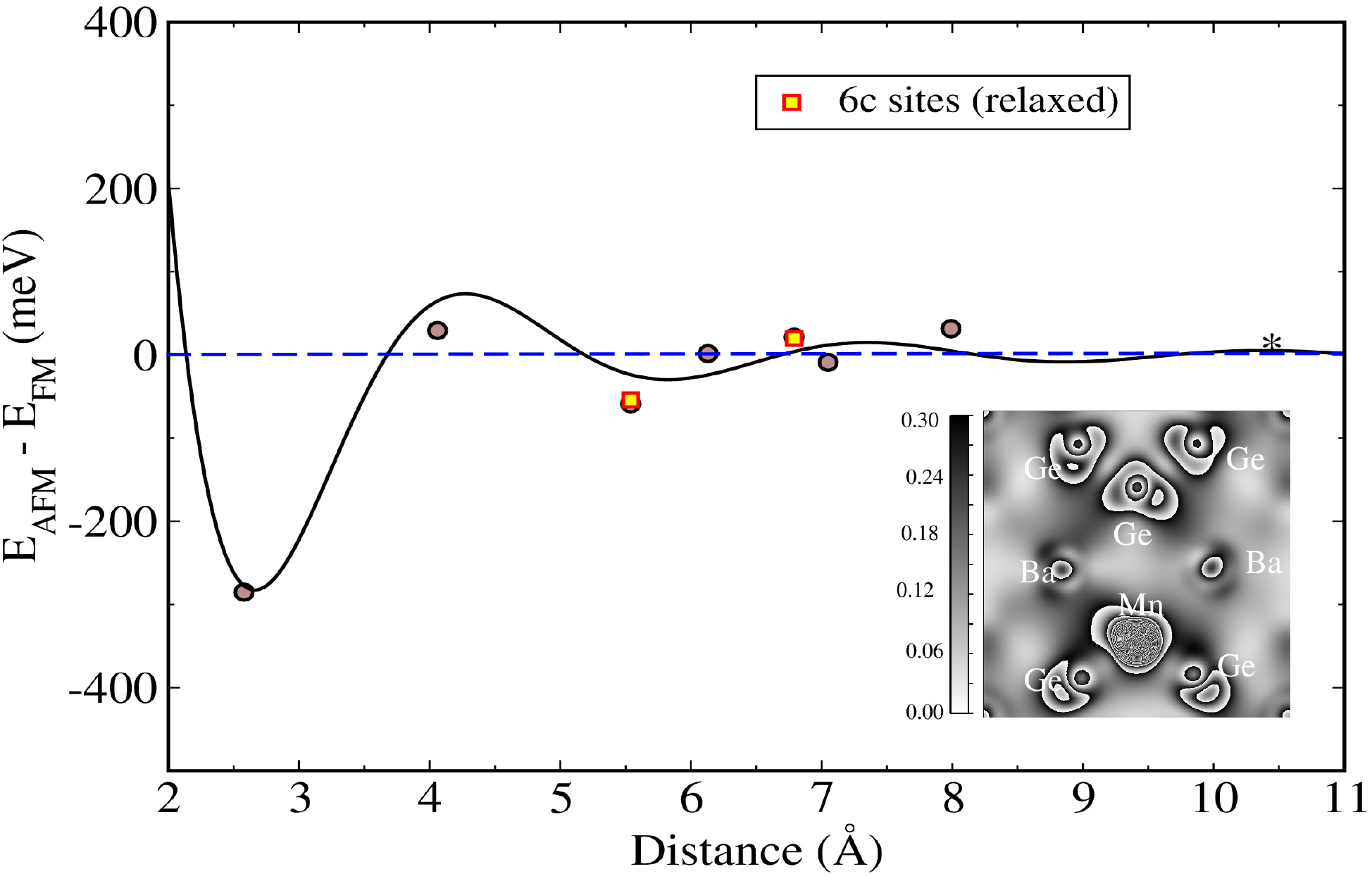}
\caption{\label{fig:exRkky}(Color online) The variation of magnetic exchange interaction strength with distance between the Mn atoms has been plotted and fitted to the RKKY expression with $k_F = 1.05$~\AA$^{-1}$. Inset shows the 2-dimensional charge density plot on a plane containing one dopant Mn atom.}
\end{figure}
 We have fitted the calculated exchange interactions with the RKKY expression for the exchange interaction strength $J(r)$ given by
\begin{eqnarray}
J(r)= {\rm Const.} \times [\sin(2k_Fr) - 2k_Fr \cos(2k_Fr)]/r^4,
\end{eqnarray}
where r is the distance between dopant magnetic atoms and $k_F$ is the Fermi wave vector corresponding to the average electron density.\cite{freemanPRL03} Fig.~\ref{fig:exRkky} shows excellent agreement of the fit with RKKY model with the fitting parameter $k_F = 1.05$~\AA$^{-1}$, confirming RKKY mechanism to be operative in Mn doped Ba$_8$Ge$_{46}$ clathrates. Further the plot of the charge density shown in  Fig.~\ref{fig:exRkky}(inset) reveal that it is delocalized in the entire system and the exchange interaction is mediated by the conduction electron cloud.

We shall now compare our calculated results with the available experimental data. In view of the large average distance (d) between Mn atoms located at the 6c sites (d=8.5$\AA$ for lattice constant a = 10.689 $\AA$ ) reported for this system \cite{kawaguchi00} we have extrapolated the RKKY plot to obtain the exchange interaction between the Mn ions located at the 6c sites other than configuration~I  and configuration~II. Fig. 4 reveal that the exchange interaction is not only ferromagnetic for configuration~II (d= 6.79 $\AA$ for theoretically calculated lattice constant a=11.08 $\AA$) but also when the separation between a pair of Mn atoms at the 6c site is d=10.36 $\AA$.
The large interval between the Mn atoms reported in the experimental work \cite{kawaguchi00} possibly refers to the latter two configurations where the exchange interaction is always ferromagnetic. The small saturation moment (0.8~$\mu_B$) obtained experimentally finds a natural explanation in such an RKKY scenario. The Mn locations are random due to competitive total energies (either FM or AFM) and the Mn atoms that prefer the AFM state do not contribute to the magnetic moment and the moment is due to a small fraction of Mn atoms that are magnetically active in the FM state. The weak exchange interaction\cite{SI} for Mn doped Ba$_8$Ge$_{46}$, also suggests a low Curie temperature in Ba$_8$Mn$_2$Ge$_{44}$, in agreement with the experiment.

In conclusion, we have studied the electronic structure of Ge$_{46}$ and Ba$_8$Ge$_{46}$ clathrates doped with Mn from {\em ab~initio} density functional calculations. We find that ferromagnetic ground state may be realized in both the compounds. The origin of ferromagnetism in Mn$_2$Ge$_{44}$ is driven by hybridization induced negative exchange splitting while it is RKKY-like for Mn doped Ba$_8$Ge$_{46}$. The origin of the two different mechanism may be traced back to the electronic structure of these systems and limit of validity of the RKKY model. The RKKY limit, $\frac{E_{x}}{E_{F}} << 1$, where $E_{x}$ is the exchange splitting of the host band and $E_{F}$ is the Fermi energy is not satisfied for the half-metallic system Mn$_2$Ge$_{44}$ due to complete spin polarization, resulting in $E_{x}> E_{F}$.\cite{mahadevan04} However incorporation of the conduction electrons in Ba$_8$Mn$_2$Ge$_{44}$  upon Ba encapsulation makes the system metallic and protects the RKKY limit. The RKKY-like scenario predicted for Ba$_8$Mn$_2$Ge$_{44}$ is also consistent with the major experimental observations for this system.

The authors acknowledge financial support from DST India (No. INT /EC /MONAMI /(28) /233513 /2008).

\begin{thebibliography}{27}
\expandafter\ifx\csname natexlab\endcsname\relax\def\natexlab#1{#1}\fi
\expandafter\ifx\csname bibnamefont\endcsname\relax
  \def\bibnamefont#1{#1}\fi
\expandafter\ifx\csname bibfnamefont\endcsname\relax
  \def\bibfnamefont#1{#1}\fi
\expandafter\ifx\csname citenamefont\endcsname\relax
  \def\citenamefont#1{#1}\fi
\expandafter\ifx\csname url\endcsname\relax
  \def\url#1{\texttt{#1}}\fi
\expandafter\ifx\csname urlprefix\endcsname\relax\def\urlprefix{URL }\fi
\providecommand{\bibinfo}[2]{#2}
\providecommand{\eprint}[2][]{\url{#2}}

\bibitem[{\citenamefont{Adams et~al.}(1994)\citenamefont{Adams, O'Keeffe,
  Demkov, Sankey, and Huang}}]{adams94}
\bibinfo{author}{\bibfnamefont{G.~B.} \bibnamefont{Adams}},
  \bibinfo{author}{\bibfnamefont{M.}~\bibnamefont{O'Keeffe}},
  \bibinfo{author}{\bibfnamefont{A.~A.} \bibnamefont{Demkov}},
  \bibinfo{author}{\bibfnamefont{O.~F.} \bibnamefont{Sankey}},
  \bibnamefont{and} \bibinfo{author}{\bibfnamefont{Y.~M.} \bibnamefont{Huang}},
  \bibinfo{journal}{Phys. Rev. B} \textbf{\bibinfo{volume}{49}},
  \bibinfo{pages}{8048} (\bibinfo{year}{1994}).

\bibitem[{\citenamefont{Moriguchi et~al.}(2000)\citenamefont{Moriguchi,
  Yonemura, Shintani, and Yamanaka}}]{moriguchi00}
\bibinfo{author}{\bibfnamefont{K.}~\bibnamefont{Moriguchi}},
  \bibinfo{author}{\bibfnamefont{M.}~\bibnamefont{Yonemura}},
  \bibinfo{author}{\bibfnamefont{A.}~\bibnamefont{Shintani}}, \bibnamefont{and}
  \bibinfo{author}{\bibfnamefont{S.}~\bibnamefont{Yamanaka}},
  \bibinfo{journal}{Phys. Rev. B} \textbf{\bibinfo{volume}{61}},
  \bibinfo{pages}{9859} (\bibinfo{year}{2000}).

\bibitem[{\citenamefont{Yuan et~al.}(2004)\citenamefont{Yuan, Grosche,
  Carrillo-Cabrera, Pacheco, Sparn, Baenitz, Schwarz, Grin, and
  Steglich}}]{yuan04}
\bibinfo{author}{\bibfnamefont{H.~Q.} \bibnamefont{Yuan}},
  \bibinfo{author}{\bibfnamefont{F.~M.} \bibnamefont{Grosche}},
  \bibinfo{author}{\bibfnamefont{W.}~\bibnamefont{Carrillo-Cabrera}},
  \bibinfo{author}{\bibfnamefont{V.}~\bibnamefont{Pacheco}},
  \bibinfo{author}{\bibfnamefont{G.}~\bibnamefont{Sparn}},
  \bibinfo{author}{\bibfnamefont{M.}~\bibnamefont{Baenitz}},
  \bibinfo{author}{\bibfnamefont{U.}~\bibnamefont{Schwarz}},
  \bibinfo{author}{\bibfnamefont{Y.}~\bibnamefont{Grin}}, \bibnamefont{and}
  \bibinfo{author}{\bibfnamefont{F.}~\bibnamefont{Steglich}},
  \bibinfo{journal}{Phys. Rev. B} \textbf{\bibinfo{volume}{70}},
  \bibinfo{pages}{174512} (\bibinfo{year}{2004}).

\bibitem[{\citenamefont{Nasir et~al.}(2009)\citenamefont{Nasir, Grytsiv,
  Melnychenko-Koblyukl, Rogl, Bauer, Lackner, Royanian, Giester, and
  Saccone}}]{nasir09}
\bibinfo{author}{\bibfnamefont{N.}~\bibnamefont{Nasir}},
  \bibinfo{author}{\bibfnamefont{A.}~\bibnamefont{Grytsiv}},
  \bibinfo{author}{\bibfnamefont{N.}~\bibnamefont{Melnychenko-Koblyukl}},
  \bibinfo{author}{\bibfnamefont{P.}~\bibnamefont{Rogl}},
  \bibinfo{author}{\bibfnamefont{E.}~\bibnamefont{Bauer}},
  \bibinfo{author}{\bibfnamefont{R.}~\bibnamefont{Lackner}},
  \bibinfo{author}{\bibfnamefont{E.}~\bibnamefont{Royanian}},
  \bibinfo{author}{\bibfnamefont{G.}~\bibnamefont{Giester}}, \bibnamefont{and}
  \bibinfo{author}{\bibfnamefont{A.}~\bibnamefont{Saccone}},
  \bibinfo{journal}{J. Phys.: Condens. Matter} \textbf{\bibinfo{volume}{21}},
  \bibinfo{pages}{385404} (\bibinfo{year}{2009}).

\bibitem[{\citenamefont{Neiner et~al.}(2007)\citenamefont{Neiner, Okamoto,
  Condron, Ramasse, Yu, Browning, and Kauzlarich}}]{neiner07}
\bibinfo{author}{\bibfnamefont{D.}~\bibnamefont{Neiner}},
  \bibinfo{author}{\bibfnamefont{N.~L.} \bibnamefont{Okamoto}},
  \bibinfo{author}{\bibfnamefont{C.~L.} \bibnamefont{Condron}},
  \bibinfo{author}{\bibfnamefont{Q.~M.} \bibnamefont{Ramasse}},
  \bibinfo{author}{\bibfnamefont{P.}~\bibnamefont{Yu}},
  \bibinfo{author}{\bibfnamefont{N.~D.} \bibnamefont{Browning}},
  \bibnamefont{and} \bibinfo{author}{\bibfnamefont{S.~M.}
  \bibnamefont{Kauzlarich}}, \bibinfo{journal}{J. Am. Chem. Soc.}
  \textbf{\bibinfo{volume}{129}}, \bibinfo{pages}{13857}
  (\bibinfo{year}{2007}).

\bibitem[{\citenamefont{Kawaguchi et~al.}(2000)\citenamefont{Kawaguchi,
  Tanigaki, and Yasukawa}}]{kawaguchi00}
\bibinfo{author}{\bibfnamefont{T.}~\bibnamefont{Kawaguchi}},
  \bibinfo{author}{\bibfnamefont{K.}~\bibnamefont{Tanigaki}}, \bibnamefont{and}
  \bibinfo{author}{\bibfnamefont{M.}~\bibnamefont{Yasukawa}},
  \bibinfo{journal}{Appl. Phys. Lett.} \textbf{\bibinfo{volume}{77}},
  \bibinfo{pages}{3438} (\bibinfo{year}{2000}).

\bibitem[{\citenamefont{Li and Ross}(2003)}]{li03}
\bibinfo{author}{\bibfnamefont{Y.}~\bibnamefont{Li}} \bibnamefont{and}
  \bibinfo{author}{\bibfnamefont{J.~H.} \bibnamefont{Ross}},
  \bibinfo{journal}{Appl. Phys. Lett.} \textbf{\bibinfo{volume}{83}},
  \bibinfo{pages}{2868} (\bibinfo{year}{2003}).

\bibitem[{\citenamefont{Woods et~al.}(2006)\citenamefont{Woods, Martin,
  Beekman, Hermann, Grandjean, Keppens, Leupold, Long, and Nolas}}]{woods06}
\bibinfo{author}{\bibfnamefont{G.~T.} \bibnamefont{Woods}},
  \bibinfo{author}{\bibfnamefont{J.}~\bibnamefont{Martin}},
  \bibinfo{author}{\bibfnamefont{M.}~\bibnamefont{Beekman}},
  \bibinfo{author}{\bibfnamefont{R.~P.} \bibnamefont{Hermann}},
  \bibinfo{author}{\bibfnamefont{F.}~\bibnamefont{Grandjean}},
  \bibinfo{author}{\bibfnamefont{V.}~\bibnamefont{Keppens}},
  \bibinfo{author}{\bibfnamefont{O.}~\bibnamefont{Leupold}},
  \bibinfo{author}{\bibfnamefont{G.~J.} \bibnamefont{Long}}, \bibnamefont{and}
  \bibinfo{author}{\bibfnamefont{G.~S.} \bibnamefont{Nolas}},
  \bibinfo{journal}{Phys. Rev. B} \textbf{\bibinfo{volume}{73}},
  \bibinfo{pages}{174403} (\bibinfo{year}{2006}).

\bibitem[{\citenamefont{Phan et~al.}(2008)\citenamefont{Phan, Woods,
  Chaturvedi, Stefanoski, Nolas, and Srikanth}}]{phan08}
\bibinfo{author}{\bibfnamefont{M.~H.} \bibnamefont{Phan}},
  \bibinfo{author}{\bibfnamefont{G.~T.} \bibnamefont{Woods}},
  \bibinfo{author}{\bibfnamefont{A.}~\bibnamefont{Chaturvedi}},
  \bibinfo{author}{\bibfnamefont{S.}~\bibnamefont{Stefanoski}},
  \bibinfo{author}{\bibfnamefont{G.~S.} \bibnamefont{Nolas}}, \bibnamefont{and}
  \bibinfo{author}{\bibfnamefont{H.}~\bibnamefont{Srikanth}},
  \bibinfo{journal}{Appl. Phys. Lett.} \textbf{\bibinfo{volume}{93}},
  \bibinfo{pages}{252505} (\bibinfo{year}{2008}).

\bibitem[{\citenamefont{Guloy et~al.}(2006)\citenamefont{Guloy, Ramlau, Tang,
  Schnelle, Baitinger, and Grin}}]{guloy06}
\bibinfo{author}{\bibfnamefont{A.~M.} \bibnamefont{Guloy}},
  \bibinfo{author}{\bibfnamefont{R.}~\bibnamefont{Ramlau}},
  \bibinfo{author}{\bibfnamefont{Z.}~\bibnamefont{Tang}},
  \bibinfo{author}{\bibfnamefont{W.}~\bibnamefont{Schnelle}},
  \bibinfo{author}{\bibfnamefont{M.}~\bibnamefont{Baitinger}},
  \bibnamefont{and} \bibinfo{author}{\bibfnamefont{Y.}~\bibnamefont{Grin}},
  \bibinfo{journal}{Nature} \textbf{\bibinfo{volume}{443}},
  \bibinfo{pages}{320} (\bibinfo{year}{2006}).

\bibitem[{\citenamefont{Zhao et~al.}(1999)\citenamefont{Zhao, Buldum, Lu, and
  Fong}}]{zhao99}
\bibinfo{author}{\bibfnamefont{J.}~\bibnamefont{Zhao}},
  \bibinfo{author}{\bibfnamefont{A.}~\bibnamefont{Buldum}},
  \bibinfo{author}{\bibfnamefont{J.~P.} \bibnamefont{Lu}}, \bibnamefont{and}
  \bibinfo{author}{\bibfnamefont{C.~Y.} \bibnamefont{Fong}},
  \bibinfo{journal}{Phys. Rev. B} \textbf{\bibinfo{volume}{60}},
  \bibinfo{pages}{14177} (\bibinfo{year}{1999}).

\bibitem[{\citenamefont{Candolfi et~al.}(2011)\citenamefont{Candolfi, Ormeci,
  Aydemir, Baitinger, Oeschler, Grin, and Steglich}}]{candolfiPRB11}
\bibinfo{author}{\bibfnamefont{C.}~\bibnamefont{Candolfi}},
  \bibinfo{author}{\bibfnamefont{A.}~\bibnamefont{Ormeci}},
  \bibinfo{author}{\bibfnamefont{U.}~\bibnamefont{Aydemir}},
  \bibinfo{author}{\bibfnamefont{M.}~\bibnamefont{Baitinger}},
  \bibinfo{author}{\bibfnamefont{N.}~\bibnamefont{Oeschler}},
  \bibinfo{author}{\bibfnamefont{Y.}~\bibnamefont{Grin}}, \bibnamefont{and}
  \bibinfo{author}{\bibfnamefont{F.}~\bibnamefont{Steglich}},
  \bibinfo{journal}{Phys. Rev. B} \textbf{\bibinfo{volume}{84}},
  \bibinfo{pages}{205118} (\bibinfo{year}{2011}).

\bibitem[{\citenamefont{Aydemir et~al.}(2010)\citenamefont{Aydemir, Candolfi,
  Borrmann, Baitinger, Ormeci, Carrillo-Cabrera, Chubilleau, Lenoir, Dauscher,
  Oeschler et~al.}}]{aydemirDT10}
\bibinfo{author}{\bibfnamefont{U.}~\bibnamefont{Aydemir}},
  \bibinfo{author}{\bibfnamefont{C.}~\bibnamefont{Candolfi}},
  \bibinfo{author}{\bibfnamefont{H.}~\bibnamefont{Borrmann}},
  \bibinfo{author}{\bibfnamefont{M.}~\bibnamefont{Baitinger}},
  \bibinfo{author}{\bibfnamefont{A.}~\bibnamefont{Ormeci}},
  \bibinfo{author}{\bibfnamefont{W.}~\bibnamefont{Carrillo-Cabrera}},
  \bibinfo{author}{\bibfnamefont{C.}~\bibnamefont{Chubilleau}},
  \bibinfo{author}{\bibfnamefont{B.}~\bibnamefont{Lenoir}},
  \bibinfo{author}{\bibfnamefont{A.}~\bibnamefont{Dauscher}},
  \bibinfo{author}{\bibfnamefont{N.}~\bibnamefont{Oeschler}},
  \bibnamefont{et~al.}, \bibinfo{journal}{Dalton Trans.}
  \textbf{\bibinfo{volume}{39}}, \bibinfo{pages}{1078} (\bibinfo{year}{2010}).

\bibitem[{\citenamefont{Yang et~al.}(2004)\citenamefont{Yang, Zhao, and
  Lu}}]{yang04}
\bibinfo{author}{\bibfnamefont{C.~K.} \bibnamefont{Yang}},
  \bibinfo{author}{\bibfnamefont{J.}~\bibnamefont{Zhao}}, \bibnamefont{and}
  \bibinfo{author}{\bibfnamefont{J.~P.} \bibnamefont{Lu}},
  \bibinfo{journal}{Phys. Rev. B} \textbf{\bibinfo{volume}{70}},
  \bibinfo{pages}{073201} (\bibinfo{year}{2004}).

\bibitem[{\citenamefont{Kresse and Furthm{\"{u}}ller}(1996)}]{vasp2}
\bibinfo{author}{\bibfnamefont{G.}~\bibnamefont{Kresse}} \bibnamefont{and}
  \bibinfo{author}{\bibfnamefont{J.}~\bibnamefont{Furthm{\"{u}}ller}},
  \bibinfo{journal}{Phys. Rev. B} \textbf{\bibinfo{volume}{54}},
  \bibinfo{pages}{11169} (\bibinfo{year}{1996}).

\bibitem[{\citenamefont{Kresse and Hafner}(1993)}]{vasp1}
\bibinfo{author}{\bibfnamefont{G.}~\bibnamefont{Kresse}} \bibnamefont{and}
  \bibinfo{author}{\bibfnamefont{J.}~\bibnamefont{Hafner}},
  \bibinfo{journal}{Phys. Rev. B} \textbf{\bibinfo{volume}{47}},
  \bibinfo{pages}{558} (\bibinfo{year}{1993}).

\bibitem[{\citenamefont{Bl{\"{o}}chl}(1994)}]{paw}
\bibinfo{author}{\bibfnamefont{P.~E.} \bibnamefont{Bl{\"{o}}chl}},
  \bibinfo{journal}{Phys. Rev. B} \textbf{\bibinfo{volume}{50}},
  \bibinfo{pages}{17953} (\bibinfo{year}{1994}).

\bibitem[{\citenamefont{Perdew et~al.}(1996)\citenamefont{Perdew, Burke, and
  Ernzerhof}}]{pbe}
\bibinfo{author}{\bibfnamefont{J.~P.} \bibnamefont{Perdew}},
  \bibinfo{author}{\bibfnamefont{K.}~\bibnamefont{Burke}}, \bibnamefont{and}
  \bibinfo{author}{\bibfnamefont{M.}~\bibnamefont{Ernzerhof}},
  \bibinfo{journal}{Phys. Rev. Lett.} \textbf{\bibinfo{volume}{77}},
  \bibinfo{pages}{3865} (\bibinfo{year}{1996}).

\bibitem[{\citenamefont{Dudarev et~al.}(1998)\citenamefont{Dudarev, Botton,
  Savrasov, Humphreys, and Sutton}}]{dudarevPRB98}
\bibinfo{author}{\bibfnamefont{S.~L.} \bibnamefont{Dudarev}},
  \bibinfo{author}{\bibfnamefont{G.~A.} \bibnamefont{Botton}},
  \bibinfo{author}{\bibfnamefont{S.~Y.} \bibnamefont{Savrasov}},
  \bibinfo{author}{\bibfnamefont{C.~J.} \bibnamefont{Humphreys}},
  \bibnamefont{and} \bibinfo{author}{\bibfnamefont{A.~P.}
  \bibnamefont{Sutton}}, \bibinfo{journal}{Phys. Rev. B}
  \textbf{\bibinfo{volume}{57}}, \bibinfo{pages}{1505} (\bibinfo{year}{1998}).

\bibitem[{\citenamefont{Dong and Sankey}(1999)}]{dong99}
\bibinfo{author}{\bibfnamefont{J.}~\bibnamefont{Dong}} \bibnamefont{and}
  \bibinfo{author}{\bibfnamefont{O.~F.} \bibnamefont{Sankey}},
  \bibinfo{journal}{J. Phys.: Condens. Matter} \textbf{\bibinfo{volume}{11}},
  \bibinfo{pages}{6129} (\bibinfo{year}{1999}).

\bibitem[{\citenamefont{Saito and Oshiyama}(1995)}]{SaitoPRB95}
\bibinfo{author}{\bibfnamefont{S.}~\bibnamefont{Saito}} \bibnamefont{and}
  \bibinfo{author}{\bibfnamefont{A.}~\bibnamefont{Oshiyama}},
  \bibinfo{journal}{Phys. Rev. B} \textbf{\bibinfo{volume}{51}},
  \bibinfo{pages}{2628} (\bibinfo{year}{1995}).

\bibitem[{SI()}]{SI}
\bibinfo{note}{See the Supplementary Information}.

\bibitem[{\citenamefont{Sarma et~al.}(2000)\citenamefont{Sarma, Mahadevan,
  Saha-Dasgupta, Ray, and Kumar}}]{sarma00}
\bibinfo{author}{\bibfnamefont{D.~D.} \bibnamefont{Sarma}},
  \bibinfo{author}{\bibfnamefont{P.}~\bibnamefont{Mahadevan}},
  \bibinfo{author}{\bibfnamefont{T.}~\bibnamefont{Saha-Dasgupta}},
  \bibinfo{author}{\bibfnamefont{S.}~\bibnamefont{Ray}}, \bibnamefont{and}
  \bibinfo{author}{\bibfnamefont{A.}~\bibnamefont{Kumar}},
  \bibinfo{journal}{Phys. Rev. Lett.} \textbf{\bibinfo{volume}{85}},
  \bibinfo{pages}{2549} (\bibinfo{year}{2000}).

\bibitem[{\citenamefont{Fang et~al.}(2001)\citenamefont{Fang, Terakura, and
  Kanamori}}]{fangPRB01}
\bibinfo{author}{\bibfnamefont{Z.}~\bibnamefont{Fang}},
  \bibinfo{author}{\bibfnamefont{K.}~\bibnamefont{Terakura}}, \bibnamefont{and}
  \bibinfo{author}{\bibfnamefont{J.}~\bibnamefont{Kanamori}},
  \bibinfo{journal}{Phys. Rev. B} \textbf{\bibinfo{volume}{63}},
  \bibinfo{pages}{180407} (\bibinfo{year}{2001}).

\bibitem[{\citenamefont{Kanamori and Terakura}(2001)}]{kanamoriJPSJ01}
\bibinfo{author}{\bibfnamefont{J.}~\bibnamefont{Kanamori}} \bibnamefont{and}
  \bibinfo{author}{\bibfnamefont{K.}~\bibnamefont{Terakura}},
  \bibinfo{journal}{J. Phys. Soc. Jpn.} \textbf{\bibinfo{volume}{70}},
  \bibinfo{pages}{1433} (\bibinfo{year}{2001}).

\bibitem[{\citenamefont{Mahadevan et~al.}(2004)\citenamefont{Mahadevan, Zunger,
  and Sarma}}]{mahadevan04}
\bibinfo{author}{\bibfnamefont{P.}~\bibnamefont{Mahadevan}},
  \bibinfo{author}{\bibfnamefont{A.}~\bibnamefont{Zunger}}, \bibnamefont{and}
  \bibinfo{author}{\bibfnamefont{D.~D.} \bibnamefont{Sarma}},
  \bibinfo{journal}{Phys. Rev. Lett.} \textbf{\bibinfo{volume}{93}},
  \bibinfo{pages}{177201} (\bibinfo{year}{2004}).

\bibitem[{\citenamefont{Zhao et~al.}(2003)\citenamefont{Zhao, Shishidou, and
  Freeman}}]{freemanPRL03}
\bibinfo{author}{\bibfnamefont{Y.-J.} \bibnamefont{Zhao}},
  \bibinfo{author}{\bibfnamefont{T.}~\bibnamefont{Shishidou}},
  \bibnamefont{and} \bibinfo{author}{\bibfnamefont{A.~J.}
  \bibnamefont{Freeman}}, \bibinfo{journal}{Phys. Rev. Lett.}
  \textbf{\bibinfo{volume}{90}}, \bibinfo{pages}{047204}
  (\bibinfo{year}{2003}).

\end{thebibliography}

\newpage

\begin{table*}[h]
{\large \bf Supplementary Information}
	\renewcommand\thetable{SI-1}
	\centering
	\caption{Energy difference between antiferromagnetic and ferromagnetic states and the magnetic moments per unit cell (in the ferromagnetic state) for Mn$_2$Ge$_{44}$ in both configurations and different values of Hubbard U parameter.}
	\begin{tabular}{|c|cc|cc|}
		\hline
		$U$~(eV) & \multicolumn{2}{c|}{Configuration~I} & \multicolumn{2}{c|}{Configuration~II} \\
		 & $E_{AFM} - E_{FM}$ & magnetic moment & $E_{AFM} - E_{FM}$ & magnetic moment \\
		 & (meV) & ($\mu_B$ per cell) & (meV) & ($\mu_B$ per cell) \\
		\hline
		0.0 & 186.3 & 6.0 & 151.1 & 6.0 \\
		2.0 & 182.5 & 6.0 & 145.2 & 6.0 \\
		4.0 & 171.6 & 6.0 & 134.3 & 6.0 \\
		5.0 & 21.5 & 7.6 & 14.4 & 7.7 \\
		6.0 & 10.3 & 8.2 & 5.2 & 8.2 \\
		7.0 & -6.1 & 8.5 & -8.5 & 8.7 \\
		\hline 
	\end{tabular}
\end{table*}

\begin{table*}[h]
	\renewcommand\thetable{SI-2}
	\centering
	\caption{Energy difference between antiferromagnetic and ferromagnetic states and the magnetic moments per unit cell (in the ferromagnetic state) for Ba$_8$Mn$_2$Ge$_{44}$ in both configurations and different values of Hubbard U parameter.}
	\begin{tabular}{|c|cc|cc|}
		\hline
		$U$~(eV) & \multicolumn{2}{c|}{Configuration~I} & \multicolumn{2}{c|}{Configuration~II} \\
		 & $E_{AFM} - E_{FM}$ & magnetic moment & $E_{AFM} - E_{FM}$ & magnetic moment \\
		 & (meV) & ($\mu_B$ per cell) & (meV) & ($\mu_B$ per cell) \\
		\hline
		0.0 & -22.8 & 6.1 & 22.7 & 6.1 \\
		4.0 & -54.2 & 6.3 & 19.9 & 6.4 \\
		5.0 & -56.4 & 7.5 & 12.2 & 7.6 \\
		7.0 & -142.0 & 9.1 & 9.3 & 9.2 \\
		\hline 
	\end{tabular}
\end{table*}
\end{document}